\documentstyle{article}

\def\be{\begin{equation}}
\def\ee{\end{equation}}
\def\bea{\begin{eqnarray}}
\def\eea{\end{eqnarray}}

\begin{document}

\pagestyle{empty}
\vskip-10pt
\hfill G\"{o}teborg ITP 99-09
\vskip-10pt
\hfill {\tt hep-th/9908107}
\begin{center}
\vskip 3truecm
{\Large\bf
Holomorphic factorization of correlation functions\\
in (4k+2)-dimensional (2k)-form gauge theory
}\\ 
\vskip 2truecm
{\large\bf M{\aa}ns Henningson, Bengt E.W. Nilsson and Per Salomonson}\\
\vskip 1truecm
{\it Institute of Theoretical  Physics\\
Chalmers University of Technology\\
S-412 96 G\"{o}teborg, Sweden}\\
\vskip 5truemm
{\tt mans@fy.chalmers.se, tfebn@fy.chalmers.se, tfeps@fy.chalmers.se}
\end{center}
\vskip 2truecm
\noindent{\bf Abstract:}
We consider a free $(2 k)$-form gauge-field on a Euclidean $(4 k + 2)$-manifold. The parameters needed to specify the action and the gauge-invariant observables take their values in spaces with natural complex structures. We show that the correlation functions can be written as a finite sum of terms, each of which is a product of a holomorphic and an anti-holomorphic factor. The holomorphic factors are naturally interpreted as correlation functions for a chiral $(2 k)$-form, i.e. a $(2 k)$-form with a self-dual $(2 k + 1)$-form field strength, after Wick rotation to a Minkowski signature. 
\vfill
\vskip4pt
\noindent{August 1999}

\eject
\newpage
\pagestyle{plain}

\section{Introduction}
Higher rank abelian gauge fields play a prominent role in string theory, although their geometrical meaning is not quite clear. Locally, such a field is given by a $p$-form $B$ modulo the gauge equivalence relation $B \sim B + \Delta B$, where $\Delta B$ is a closed $p$-form with integer periods. When $p = 1$, this is just an ordinary abelian gauge field with its usual gauge invariance. 

On a $d$-manifold $M$, a natural class of gauge invariant observables can be written in the form $\exp 2 \pi i \int_M B \wedge P$, where $P$ is a closed $(d - p)$-form with integer periods. For example, if $P$ is the Poincar\'e dual of some $p$-cycle $\Sigma$ in $M$, we get the `Wilson volume' observable $\exp 2 \pi i \int_\Sigma B$ associated with $\Sigma$. A product of such observables is again of the same form with the resulting parameter $P$ being the sum of the $P$'s of the different factors. To compute the expectation value of such an observable, it is in fact sufficient to consider the case when $P$ is trivial in cohomology, i.e. when $P = 2 d Q$ for some $(d - p - 1)$-form $Q$. (The factor of $2$ is inserted for later convenience.) For example, $P$ could be twice the Poincar\'e dual of a homologically trivial $p$-cycle $\Sigma$ in $M$. The observable is then of the form 
\be\label{WQ}
W (Q) = \exp 4 \pi i \int_M H \wedge Q , 
\ee
where $H = d B$ is the gauge invariant $(p + 1)$-form field strength of $B$. The reason that we do not need to consider cohomologically non-trivial $P$ is that in this case, the $B$-field would contain a closed mode Poincar\'e dual to $P$. Furthermore, this mode does not appear in the action that we will discuss shortly. Because of the gauge invariance, the coefficient of this mode takes its values on a circle. The expectation value of the observable then vanishes, because its phase averages to zero when we integrate over this mode. This phenomenon is well known for a compact scalar in two dimensions, where it corresponds to conservation of discrete momentum (in a string theory compactification).

If $M$ is endowed with a metric $G$, it is natural to consider the generalized Maxwell action
\be\label{action}
S = - \frac{2 \pi}{g^2} \int_M H \wedge {}^* H ,
\ee
where ${}^*$ denotes the Hodge duality operator that maps a $(p + 1)$-form to a $(d - p - 1)$-form, and $g$ is a coupling constant. The classical equation of motion that follows from this action is
\be\label{EOM}
d {}^* H = 0 ,
\ee
i.e. a field strength $H$ that fulfills the classical equation of motion is harmonic. We can now calculate the (unnormalized) expectation value of the observable $W (Q)$ as
\be\label{pathintegral}
\left< W (Q) \right> = \int {\cal D} B \, W (Q) e^{- S} ,
\ee
where the functional integral is over gauge inequivalent field configurations $B$. 

In this paper, we will be concerned with the case when $d = 4 k + 2$ and $p = 2 k$ for some integer $k$. A priori, we are interested in the case when the metric $G$ has a Minkowski signature, which means that the Hodge duality operator ${}^*$ obeys ${}^* {}^* = 1$, so that its eigenvalues are $\pm 1$. We now say that a gauge field $B$ is chiral if its field strength $H = d B$ obeys the self-duality equation
\be\label{selfdual}
H = + {}^* H .
\ee
From this first order differential equation in fact follows the second-order equation of motion (\ref{EOM}). Similarly, an anti-chiral gauge field has an anti self-dual field strength. Chiral gauge fields have many important applications in string theory. For example, in $d = 2$ dimensions, chiral zero-form (scalars) appear on the world-sheet of heterotic string theory. In $d = 6$ dimensions, the world-volume theory of the $M$-theory five-brane contains a chiral two-form. Finally, in $d = 10$ dimensions, type $II B$ supergravity contains a chiral four-form. In $d = 2$ dimensions, there is a well-known interacting generalization of the theory, namely the chiral Wess-Zumino-Witten model. In $d = 6$ dimensions, the tensionless string theories generalize the theory of a free $(0, 2)$ tensor multiplet (which contains a chiral two-form). In $d = 10$ dimensions, no interacting generalization is known.

One can show that the self-duality equation (\ref{selfdual}) cannot follow from any covariant Lagrangian in the usual sense. It is therefore a bit subtle to define the quantum theory of a chiral gauge field \cite{Witten96}. The main idea is as follows: One starts with the non-chiral theory with action (\ref{action}). The field strength $H$ then contains both a self-dual and an anti self-dual part. However, since the theory is free, i.e. the action is a bilinear in the gauge field $B$, one would expect that the chiral and anti-chiral parts are decoupled from each other. Furthermore, observables of the form (\ref{WQ}) are exponentials of linear expressions in $B$. One would therefore expect that the correlation function (\ref{pathintegral}) can be written as the product of the correlation functions pertaining to a chiral gauge field and an anti-chiral gauge field. (We will explain the criterion for how to perform this factorization in the next section.) On a manifold $M$ of non-vanishing middle dimensional Betti number $b_{2 k + 1} = \dim H^{2 k + 1} (M , {\bf R})$, the story is in fact more complicated. It will turn out that there are then $2^{b_{2 k + 1}}$ different candidate chiral and anti-chiral correlation functions. Assembling these into vectors $\left< W (Q) \right>_+$ and $\left< W (Q) \right>_-$ respectively, one can write the non-chiral correlation function as
\be\label{factorization}
\left< W (Q) \right> = \left< W (Q) \right>_+ \cdot \left< W (Q) \right>_- ,
\ee
where the raised dot denotes the vector scalar product. To define the correlation function of a chiral gauge-field, one would thus need some extra discrete data to pick out one of the terms in (\ref{factorization}). In applications, this choice is in fact often dictated by the physical context \cite{Witten96}. (On a torus, it turns out that there is a canonical choice which is invariant under the $SL(4 k + 2, {\bf Z})$ mapping class group. The corresponding chiral partition function was calculated in the case of a flat metric in \cite{Dolan-Nappi}.)

In this paper, we will carry out the procedure outlined in \cite{Witten96} in detail and derive (\ref{factorization}) explicitly. We will consider the exact quantum correlation function for a general observable of the form (\ref{WQ}). (The discussion in \cite{Witten96} focuses on the partition function and the contributions to the functional integral (\ref{pathintegral}) from field configurations that solve the classical equations of motion (\ref{EOM}). In that paper, the theory is coupled to a background $(2 k + 1)$-form, that is put to zero in this paper.) In the next section, we will describe how to separate a non-chiral correlation function into its chiral and anti-chiral parts. In fact, the non-chiral correlation function also contains certain `anomalous' factors that have to be discarded in order for the factorization (\ref{factorization}) to work. In section~3, we will evaluate the contributions due to classical field configurations, and in the last section we will consider the quantum fluctuations. 
  
\section{Holomorphic factorization}
\subsection{The complex structure}
We thus consider the case of a $(2 k + 1)$-form gauge field $B$ in a $(4 k + 2)$-dimensional space $M$. The coupling constant $g$ in (\ref{action}) is then dimensionless, so that the action is scale invariant and in fact only depends on the conformal equivalence class $[G]$ of the metric $G$ on $M$. (Our treatment in this paper is purely formal, and we will thus neglect conformal anomalies such as the one discussed in \cite{Henningson-Skenderis}.) Actually, $[G]$ only appears in (\ref{action}) through the Hodge duality operator ${}^*$ that maps the space
\be\label{Omega}
\Omega = \{ (2 k + 1)-{\rm forms \;\; on \;\;} M \} ,
\ee
of which the field strength $H$ is an element, to itself. We note that the parameter $Q$ in the observable (\ref{WQ}) is also an element of $\Omega$.

To describe how to separate the non-chiral correlation function (\ref{pathintegral}) into its chiral and anti-chiral parts, we will actually have to assume that the metric $G$ (and thus the conformal structure $[G]$) of $M$ has a Euclidean signature. The Hodge duality operator ${}^*$ then obeys ${}^* {}^* = -1$, and can thus be thought of as defining a complex structure $J$ on the space $\Omega$ defined in (\ref{Omega}). The moduli space ${\cal J}$ of all complex structures on $\Omega$ itself carries a natural complex structure. It is easy to show that the map described above from the moduli space ${\cal G}$ of conformal structures on $M$ to ${\cal J}$ is injective, but not surjective. As far as we know, the image of this map, which can thus be identified with ${\cal G}$, does not carry a canonical complex structure.

Here it is in place to make some comments on the much-studied case $d = 2$, which has some additional simplifying features. In this dimension, the space $\Omega$ is simply the cotangent space, and it is well-known that a choice of a Euclidean conformal structure $[G]$ on $M$ induces a complex structure not only on $\Omega$ but on $M$ itself. Furthermore, the moduli space ${\cal J}$ of such complex structures is finite-dimensional. However, these special properties are not necessary for our discussion, and in the sequel we will consider the general case of $d = 4 k + 2$ for arbitrary $k$.

To describe the complex structure $J$ on the space $\Omega$ of $(2 k + 1)$-forms on $M$ more concretely, we begin by noting that $\Omega$ carries a natural symplectic structure. The symplectic product is simply given by the wedge product followed by integration over $M$. We can choose a basis $(E_A)_i$ and $(E_B)_i$ of $\Omega$ that is symplectic in the sense that
\bea\label{symplectic}
\int_M (E_A)_i \wedge (E_A)_j & = & \int_M (E_B)_i \wedge (E_B)_j = 0 \cr
\int_M (E_B)_i \wedge (E_A)_j & = & - \int_M (E_A)_i \wedge (E_B)_j = \delta_{i j} .
\eea
Henceforth we will use a more compact notation where the basis elements $(E_A)_i$ and $(E_B)_j$ are assembled into two (infinite dimensional) column vectors $E_A$ and $E_B$. The above conditions can then be written as
\bea
\int_M E_A \wedge E_A^t & = & \int_M E_B \wedge E_B^t = 0 \cr
\int_M E_B \wedge E_A^t & = & - \int_M E_A \wedge E_B^t = 1 ,
\eea
where the $1$ on the right hand side of the second equation denotes the unit matrix and ${}^t$ denotes the transpose. An analogous notation will be employed for other spaces throughout this paper. An element of $\Omega$, such as the field strength $H$, can be expanded in this basis as
\be\label{Hexpansion}
H = H_A^t E_A + H_B^t E_B
\ee
with some coefficient vectors $H_A$ and $H_B$. The same of course applies to the $(2 k + 1)$-form $Q$ that enters in the observable (\ref{WQ}).

Sofar, everything has been independent of the Hodge duality operator ${}^*$. For a generic choice of ${}^*$, the entries of any two of the vectors $E_A$, $E_B$, ${}^* E_A$, and ${}^* E_B$ span $\Omega$. We can for example expand $E_B$ as a linear combination of $E_A$ and ${}^* E_A$, i.e.
\be
E_B = X E_A + Y {}^* E_A .
\ee
One can show that the real coefficient matrices $X$ and $Y$ are symmetric, and with a suitable choice of $E_A$ and $E_B$ (subject to the relations above), $Y$ is positive definite. The Hodge duality operator ${}^*$, and thus the complex structure $J$ on $\Omega$, is completely determined by $X$ and $Y$, which thus can be regarded as coordinates on the moduli space ${\cal J}$ of complex structures on $\Omega$. The complex structure on ${\cal J}$ can be described by declaring that the complex linear combination
\be
Z = X + i Y
\ee
is a holomorphic coordinate on ${\cal J}$. According to the above, it fulfills the conditions
\bea\label{RiemannIII}
Z & = & Z^t \cr
{\rm Im \;} Z & > & 0 .
\eea

It is often convenient to change basis for $\Omega$ from $E_A$ and $E_B$ to a holomorphic and anti-holomorphic basis $E_+$ and $E_-$ defined as
\bea
E_+ & = & (Z - \bar{Z})^{-1} (E_B - \bar{Z} E_A) \cr
E_- & = & - (Z - \bar{Z})^{-1} (E_B - Z E_A) .
\eea
which fulfill ${}^* E_+ = + i E_+$ and ${}^* E_- = - i E_-$. We can then expand for example the field strength $H$ (or the parameter $Q$) as
\bea
H = H_+^t E_+ + H_-^t E_- ,
\eea
where the coefficient vectors $H_+$ and $H_-$ are related to $H_A$ and $H_B$ as
\bea
H_+ & = & H_A + Z H_B \cr
H_- & = & H_A + \bar{Z} H_B .
\eea

\subsection{Holomorphicity of $\left< W (Q) \right>_+$}
We will now describe a criterion for distinguishing between the contributions to the correlation functions from a chiral and an anti-chiral gauge field.

With the notation introduce above, we can write the action (\ref{action}) as
\be
S = \frac{4 \pi i}{g^2} (H_A + Z H_B)^t (Z - \bar{Z})^{-1} (H_A + \bar{Z} H_B) .
\ee
Remembering that $H_A$ and $H_B$ are independent of $\bar{Z}$, we find that under an infinitesimal anti-holomorphic variation $\delta \bar{Z}$,
\be
\delta S = \frac{4 \pi i}{g^2} (H_A + Z H_B)^t (Z - \bar{Z})^{-1} \delta \bar{Z} (Z - \bar{Z})^{-1} (H_A + Z H_B) .
\ee
Since the observable (\ref{WQ}) is independent of $\bar{Z}$ we then get, by differentiating under the integration sign in (\ref{pathintegral}) that $\delta \left< W (Q) \right> = - \left< W (Q) \delta S \right>$. For a chiral gauge field, $H_+ = H_A + Z H_B$ is zero, and the above expression for $\delta S$ would thus vanish identically. It is then natural to conjecture that the correlation function for such a theory should be holomorphic, i.e.
\be
\frac{\delta}{\delta \bar{Z}} \left< W (Q) \right>_+ = 0 .
\ee
Similarly, the correlation function $\left< W(Q) \right>_-$ of an anti-chiral gauge field should be anti-holomorphic. The requirement that the non-chiral correlation function $\left< W(Q) \right>$ in (\ref{factorization}) can be holomorphically factorized in this manner is a non-trivial constraint, that we will verify in the remainder of this paper. (In fact, we will see that this is only true if certain non-holomorphic `anomalous' factors are discarded.) In $d = 2$ dimensions, such a holomorphic factorization is an important and well-known feature of many conformal field theories, e.g. Wess-Zumino-Witten models and cosets thereof \cite{Witten92}.

\subsection{Classical and quantum contributions}
We can decompose the space $\Omega$ of $(2 k + 1)$-forms as
\be
\Omega = \Omega^0 \oplus \Omega^\prime ,
\ee
where $\Omega^0$ is the (finite-dimensional) subspace of harmonic forms, and $\Omega^\prime$ its orthogonal complement (with respect to the symplectic metric). The Hodge duality operator ${}^*$ respects this decomposition. The field strength $H$ and the parameter $Q$ of the observable (\ref{WQ}) can accordingly be decomposed as
\bea\label{decomposition}
H & = & H^0 + H^\prime \cr
Q & = & Q^0 + Q^\prime .
\eea
By the Hodge-de Rham theorem, $H^0$ is then the unique harmonic representative of the cohomology class $[H]$ of $H$, whereas $H^\prime = d B^\prime$ for some globally defined $2 k$-form $B^\prime$. In view of (\ref{EOM}), it is natural to regard $H^\prime$ as a quantum fluctuation around a classical field configuration $H^0$. Inserting (\ref{decomposition}) into the formulas (\ref{action}) and (\ref{WQ}), we then get $S = S^0 + S^\prime$ and $W (Q) = W (Q^0) W (Q^\prime)$, where
\bea\label{action2}
S^0 & = & - \frac{2 \pi}{g^2} \int_M H^0 \wedge {}^* H^0 \cr
S^\prime  & = & - \frac{2 \pi}{g^2} \int_M H^\prime \wedge {}^* H^\prime
\eea
and
\bea\label{WQ2}
W (Q^0) & = & \exp 4 \pi i \int_M H^0 \wedge Q^0 \cr
W (Q^\prime) & = & \exp 4 \pi i \int_M H^\prime \wedge Q^\prime .
\eea
In this way, we get $\left< W (Q) \right> = \left< W (Q^0) \right> \left< W (Q^\prime) \right>$, where the classical and quantum contributions are given by
\be\label{pathintegralc}
\left< W (Q^0) \right> = \sum_{H^0} W (Q^0) e^{-S^0}
\ee
and
\be\label{pathintegralq}
\left< W (Q^\prime) \right> = \int {\cal D} B^\prime W (Q^\prime) e^{-S^\prime}
\ee
respectively. The sum in the classical part is over harmonic $(2 k + 1)$-forms $H^0$ with integer periods, and the functional integral in the quantum part is over globally defined $(2 k)$-forms $B^\prime$ modulo closed forms with integer periods. 

\section{The classical contribution}
\subsection{The Poisson resummation}
The classical field strength $H^0$ belongs to the lattice $\Gamma \subset \Omega^0$ of harmonic $(2 k + 1)$-forms with integer periods. We now introduce a basis $E_A^0$ and $E_B^0$ for this lattice, which is symplectic in the same sense as $E_A$ and $E_B$ in (\ref{symplectic}). A difference is of course that $E_A^0$ and $E_B^0$ are finite dimensional vectors with $\frac{1}{2} b_{2 k + 1} = \frac{1}{2} \rm dim H^{2 k + 1} (M, {\bf R})$ entries each. By a reasoning exactly analogous to that in the paragraph following (\ref{Hexpansion}), we see that the Hodge duality operator ${}^*$ induces a conformal structure $J^0$ parametrized by $Z^0 = X^0 + i Y^0$ on the space $\Omega^0$ of harmonic $(2 k + 1)$-forms.  We can then also introduce the corresponding holomorphic and anti-holomorphic basis $E_+^0$ and $E_-^0$ of $\Omega^0$.

The classical field strength $H^0$ can now be expanded as
\bea
H^0 & = & H_A^{0 t} E_A^0 + H_B^{0 t} E_B^0 \cr
& = & H_+^{0 t} E_+^0 + H_-^{0 t} E_-^0 ,
\eea
where the coefficient vectors $H_A^0$ and $H_B^0$ have integer entries. Similarly, we expand the harmonic part $Q^0$ of the observable parameter as
\bea
Q^0 & = & Q_A^{0 t} E_A^0 + Q_B^{0 t} E_B^0 \cr
& = & Q_+^{0 t} E_+^0 + Q_-^{0 t} E_-^0 .
\eea
 Inserting these expansions in (\ref{action2}) - (\ref{pathintegralc}) and fixing the coupling constant $g$ at the particular value $g = 1$, we get
\bea
\left< W (Q^0) \right> & = & \sum_{H_A^0, H_B^0} \exp \Biggl[ - 2 \pi \Bigl( H_A^{0 t} (Y^0)^{-1} H_A^0 + 2 H_A^{0 t} (Y^0)^{-1} X^0 H_B^0 + H_B^{0 t} (X^0 (Y^0)^{-1} X^0 + Y^0) H_B^0 \Bigr) \cr
& & \;\;\;\;\;\;\;\;\;\;\;\;\;\;\;\;\;\; + 4 \pi i \left( H_B^{0 t} Q_A^0 - H_A^{0 t} Q_B^0 \right) \Biggr] .
\eea 
We now perform a Poisson resummation, replacing the sum over $H_A^0$ with a sum over a vector $m$ with integer entries. Renaming the vector $H_B^0$ as $n$, we then get
\bea
\left< W (Q^0) \right> & = & \sqrt{\det \frac{1}{2} Y^0} \sum_{n, m} \exp - 2 \pi \Bigl( n^t Y^0 n - i m^t X^0 n + \frac{1}{4} m^t Y^0 m \cr
& & \;\;\;\;\;\;\;\;\;\;\;\;\;\;\;\;\;\;\;\;\;\;\;\;\;\;\;\;\;\;\;\;\;\;\;\;\;\;\;\;\;\;\; - 2 i n^t Q_A^0 - 2 i n^t X^0 Q_B^0 + m^t Y^0 Q_B^0 + Q_B^{0 t} Y^0 Q_B^0 \Bigr) ,
\eea
or, in terms of complex variables,
\bea\label{classical}
\left< W (Q^0) \right> & = & \sqrt{\det \frac{1}{4 i} (Z^0 - \bar{Z^0})} \sum_{n, m} \exp i \pi \Bigl( (m/2 + n)^t Z^0 (m/2 + n) - (m/2 - n)^t \bar{Z^0} (m/2 - n) \cr
& & \;\;\;\;\;\;\;\; + 2 (m/2 + n)^t Q_+^0 - 2 (m/2 - n)^t Q_-^0 + (Q_-^0 - Q_+^0)^t (Z^0 - \bar{Z^0})^{-1} (Q_-^0 - Q_+^0) \Bigr) .
\eea

\subsection{A digression on theta-functions}
The coefficients $Q_A^0$ and $Q_B^0$ can be regarded as real coordinates on the space $\Omega^0$ of harmonic $(2 k + 1)$-forms. We will be interested in the torus $\Omega^0 / \Gamma$ (the intermediate Jacobian of the manifold $M$), where $\Gamma$ is the lattice of forms with integer periods. This means that we should identify
\bea
Q_A^0 & \sim & Q_A^0 + \lambda_A^0 \cr
Q_B^0 & \sim & Q_B^0 + \lambda_B^0
\eea
for any vectors $\lambda_A^0$ and $\lambda_B^0$ with integer entries. The natural complex structure on $\Omega^0$ means that $\Omega^0 / \Gamma$ can be regarded as the complex torus ${\bf C}^{\frac{1}{2} b_{2 k + 1}} / \Gamma$ with a holomorphic coordinate $Q_+^0$ subject to the identifications
\bea
Q_+^0 & \sim & Q_+^0 + \lambda_A^0 \cr
Q_+^0 & \sim & Q_+^0 + Z^0 \lambda_B^0 .
\eea
The parameter $Z^0$ obeys conditions analogous to (\ref{RiemannIII}), which implies that $\Omega^0 / \Gamma$ is an Abelian variety \cite{Griffiths-Harris}. It is endowed with a (principal) polarization given by the cohomology class $[\omega]$ of the symplectic form $\omega$ corresponding to the natural symplectic structure on $\Omega^0$.

Consider now a unitary line bundle ${\cal L}$ over $\Omega^0 / \Gamma$ with a connection whose curvature equals $\omega$. Since $\omega$ is of type $(1, 1)$, ${\cal L}$ is in fact a holomorphic line bundle. As a complex line bundle, ${\cal L}$ is determined by its Chern class $c_1 ({\cal L}) = [\omega]$, but there is a finer classification of holomorphic line bundles \cite{Griffiths-Harris}. Indeed, the kernel of the Chern class $c_1$ is isomorphic to the complex torus $H^1 (\Omega^0 / \Gamma, {\cal O}) / H^1 (\Omega^0 / \Gamma, {\bf Z})$ (the Picard variety or the dual Abelian variety), where ${\cal O}$ is the sheaf of holomorphic functions. The set of holomorphic line bundles ${\cal L}$ with $c_1 ({\cal L}) = [\omega]$ is thus also a torus, which we can parametrize by $\alpha$ and $\beta$ that are vectors of dimension $\frac{1}{2} b_{2 k + 1}$ with entries with values in ${\bf R} / {\bf Z}$. We denote the corresponding line bundle as ${\cal L} [{}^\alpha_\beta]$.

It follows from an index theorem together with the Kodaira vanishing theorem that ${\cal L} [{}^\alpha_\beta]$ has a unique holomorphic section (up to a multiplicative constant). In a certain trivialization of ${\cal L} [{}^\alpha_\beta]$, this is given by the Jacobi theta-function
\be\label{theta}
\theta \left[ {}^\alpha_\beta \right] (Z^0 | Q_+^0) = \sum_k \exp i \pi \left( (k + \alpha)^t Z^0 (k + \alpha) + 2 (k + \alpha)^t (Q_+^0 + \beta) \right) ,
\ee
where the sum runs over vectors $k$ of dimension $\frac{1}{2} b_{2 k + 1}$ with integer entries. The theta-function obeys the quasi-periodicity conditions
\bea
\theta \left[ {}^\alpha_\beta \right] (Z^0 | Q_+^0 + \lambda_A^0) / \theta \left[ {}^\alpha_\beta \right] (Z^0 | Q_+^0) & = & \exp 2 \pi i \alpha^t \lambda_A^0 \cr
\theta \left[ {}^\alpha_\beta \right] (Z^0 | Q_+^0 + Z^0 \lambda_B^0) / \theta \left[ {}^\alpha_\beta \right] (Z^0 | Q_+^0) & = & \exp i \pi \left( - \lambda_B^0 Z^0 \lambda_B^0 - 2 \lambda_B^{0 t} (Q_+^0 + \beta) \right) ,
\eea
so in this trivialization, the transition functions are holomorphic ${\bf C}^*$-valued functions. We will use a different trivialization, though, where the unique holomorphic section of ${\cal L} [{}^\alpha_\beta]$ is given by the function
\be\label{Theta}
\Theta \left[ {}^\alpha_\beta \right] (Z^0| Q_+^0, Q_-^0) = \exp i \pi Q_+^{0 t} Q_B^0 \, \theta \left[ {}^\alpha_\beta \right] (Z^0 | Q_+^0) .
\ee
Its quasi-periodicity properties are
\bea
\Theta \left[ {}^\alpha_\beta \right] (Z^0 | Q_+^0 + \lambda_A^0, Q_-^0 + \lambda_A^0) / \Theta \left[ {}^\alpha_\beta \right] (Z^0 | Q_+^0, Q_-^0) & = & \exp i \pi \lambda_A^{0 t} (Q_B^0 + 2 \alpha) \cr
\Theta \left[ {}^\alpha_\beta \right] (Z^0 | Q_+^0 + Z^0 \lambda_B^0, Q_-^0 + \bar{Z^0} \lambda_B^0) / \Theta \left[ {}^\alpha_\beta \right] (Z^0 | Q_+^0, Q_-^0) & = & \exp -i \pi \lambda_B^{0 t} (Q_A^0 + 2 \beta) ,
\eea
so in this trivialization, the transition functions are $U(1)$-valued. Finally we note that $\Theta [{}^\alpha_\beta] (Z^0| Q_+^0, Q_-^0)$ is holomorphic in the sense that it is annihilated by $\frac{\delta}{\delta \bar{Z^0}}$ and by the covariant derivative $\frac{D}{D Q_-^0} = \frac{\delta}{\delta Q_-^0} + i \pi  Q_+^{0 t} (Z^0 - \bar{Z^0})^{-1}$.

\subsection{Holomorphic factorization}
We can now rewrite the expression (\ref{classical}) for $\left< W (Q^0) \right>$ in terms of the functions $\Theta [{}^\alpha_\beta] (Z^0| Q_+^0, Q_-^0)$ introduced in (\ref{Theta}):
\be\label{classical2}
\left< W (Q^0) \right> = 2^{-\frac{1}{2} b_{2 k + 1}} \sqrt{\det \frac{1}{4 i} (Z^0 - \bar{Z^0})} \sum_{\alpha \beta} \Theta \left[ {}^\alpha_\beta \right] (Z^0| Q_+^0, Q_-^0) \overline{\Theta \left[ {}^\alpha_\beta \right] (Z^0| Q_+^0, Q_-^0)} ,
\ee
where the sum over $\alpha$ and $\beta$ runs over vectors with entries $0$ or $\frac{1}{2}$. This formula can be verified as follows: We first perform the sum over $\beta$, which gives rise to the factor $2^{\frac{1}{2} b_{2 k + 1}}$ and restricts the sums over $k$ in the definition (\ref{theta}) of $\theta [{}^\alpha_\beta] (Z^0 | Q_+^0)$ and the corresponding vector $l$ in the complex conjugate to values such that $k - l$ is a vector with even entries. The sums over $k$ and $l$ can then be exchanged to sums over the vectors $m^\prime$ and $n$ with integer entries defined through the relations $k = m^\prime + n$ and $l = m^\prime - n$. Finally, the sums over $m^\prime$ and $\alpha$ can be exchanged to a sum over the vector $m = 2 (m^\prime + \alpha)$, which yields (\ref{classical}).

We must now pick one of the terms in (\ref{classical2}), i.e. a particular value of $\alpha$ and $\beta$. The chiral part of the classical correlation function is then
\be
\left< W(Q^0) \right>_+ = \Theta \left[ {}^\alpha_\beta \right] (Z^0| Q_+^0, Q_-^0) ,
\ee
and the anti-chiral part is its complex conjugate. This leaves the `anomalous' prefactor 
\be\label{anomc}
\left<W (Q^0) \right>_{anom} = 2^{-\frac{1}{2} b_{2 k + 1}} \sqrt{\det \frac{1}{4 i} (Z^0 - \bar{Z^0})} 
\ee
in (\ref{classical2}), which is neither holomorphic nor anti-holomorphic.   

\section{The quantum contribution}
The quantum part of the gauge field is given by a globally defined $(2 k)$-form $B^\prime$ modulo closed forms. Introducing a basis $F$ in the space of such forms, we can then expand $B^\prime$ as
\be
B^\prime = b^t F
\ee
for some vector $b$ of coefficients. The exterior derivative $d$ acting on $F$ gives a vector $d F$ of exact $(2 k + 1)$-forms, which can be expressed in terms of the symplectic basis $E_A$ and $E_B$ or the holomorphic basis $E_+$ and $E_-$:
\bea
d F & = & = \Pi_A^t E_A + \Pi_B^t E_B \cr
& = & \Pi_+^t E_+ + \Pi_-^t E_- 
\eea  
for some matrices $\Pi_A$, $\Pi_B$, $\Pi_+$ and $\Pi_-$ of coefficients. The quantum part $H^\prime = d B^\prime$ of the field strength is then
\be
H^\prime = b^t \left(\Pi_A^t E_A + \Pi_B^t E_B\right) ,
\ee
and the quantum part of the action (\ref{action2}) is
\be
S^\prime = \frac{4 \pi i}{g^2} b^t K b ,
\ee
where the symmetric matrix $K$ is given by
\be\label{K}
K = \Pi_+^t \left(Z - \bar{Z} \right)^{-1} \Pi_- .
\ee
Expanding the quantum part $Q^\prime$ of $Q$ in (\ref{decomposition}) as
\be
Q^\prime = Q_+^t E_+ + Q_-^t E_-
\ee
we get for the quantum part of the observable (\ref{WQ2})
\be
W (Q^\prime) = \exp 4 \pi i b^t J ,
\ee
where
\be
J = \left(\Pi_+^t (Z - \bar{Z})^{-1} Q_- - \Pi_-^t (Z - \bar{Z})^{-1} Q_+ \right) .
\ee
We can now perform the Gaussian path-integral over the $B^\prime$-field in (\ref{pathintegralq}), i.e. over the coefficients $b$, with the result
\be
\left<W (Q^\prime) \right>  \sim \left(\det K \right)^{-1/2} \exp g^2 \pi i J^t K^{-1} J .
\ee
For this formula to make sense, the matrix $K$ in (\ref{K}) must be invertible. If we furthermore assume that the matrices $\Pi_+$ and $\Pi_+$ are also invertible, we can simplify further to get
\bea
\left<W (Q^\prime) \right>  & \sim & \left(\det \Pi_+^t (Z - \bar{Z})^{-1} \Pi_- \right)^{-1/2} \cr
& & \exp i \pi \left(Q_+^t (Z - \bar{Z})^{-1} \Pi_- \Pi_+^{-1} Q_+ - 2 Q_-^t (Z - \bar{Z})^{-1} Q_+ + Q_-^t (Z - \bar{Z})^{-1} \Pi_+ \Pi_-^{-1} Q_- \right) ,
\eea
where we have also put $g = 1$. Finally, we write the result in factorized form as
\be
\left<W (Q^\prime) \right> = \left<W (Q^\prime) \right>_{anom} \left<W (Q^\prime) \right>_+ \left<W (Q^\prime) \right>_- ,
\ee
where the chiral part is given by the holomorphic expression
\be
\left<W (Q^\prime) \right>_+ = \left(\det \Pi_+ \right)^{-1/2} \exp (-i \pi) Q_+^t \Pi_B \Pi_+^{-1} Q_+
\ee
and the anti-chiral part by its complex conjugate whereas the `anomalous' prefactor is
\be\label{anomq}
\left<W (Q^\prime) \right>_{anom} = \left(\det (Z - \bar{Z}) \right)^{1/2} \exp i \pi (Q_+^t - Q_-^t) (Z - \bar{Z})^{-1} (Q_+ - Q_-) .
\ee
Putting everything together, the final answer for the chiral correlation function is thus
\be
\left<W (Q) \right>_+ = \left(\det \Pi_+ \right)^{-1/2} \exp ( -i \pi) Q_+^t \Pi_B \Pi_+^{-1} Q_+ \, \Theta \left[{}^\alpha_\beta \right] (Z^0 | Q^0_+, Q^0_-) ,
\ee
where $\Theta \left[{}^\alpha_\beta \right] (Z^0 | Q^0_+, Q^0_-)$ is defined by (\ref{Theta}) and (\ref{theta}).

It remains to discuss the non-holomorphic factors (\ref{anomc}) and (\ref{anomq}). These must be discarded in order for the non-chiral correlation function to admit a holomorphic factorization. To make sense of the infinite-dimensional determinant in (\ref{anomq}), we can employ for example $\zeta$-function regularization. On a flat $(4 k + 2)$-torus (generalizing the two-dimensional calculation in \cite{Polchinski}), it turns out this determinant cancels the finite-dimensional determinant in (\ref{anomc}). It would be interesting to know if this is true also in more general situations. The remaining factor in (\ref{anomq}) is in some sense a `contact term' since it does not involve the `derivative' matrices $\Pi_+$ and $\Pi_-$.

This research was supported by the Swedish Natural Science Research Council (NFR).


\begin{thebibliography}{99}

\bibitem{Henningson-Skenderis}
M. Henningson and K. Skenderis, `Weyl anomaly for Wilson surfaces', {\it JHEP} {\bf 9906} (1999) 012, {\tt hep-th/9905163}.

\bibitem{Witten96}
E. Witten, `Five-brane effective action in $M$-theory', {\it J. Geom. Phys.} {\bf 22} (1997) 103, {\tt hep-th/9610234}.

\bibitem{Dolan-Nappi}
L. Dolan and C. Nappi, `A modular invariant partition function for the five-brane', {\it Nucl. Phys.} {\bf B530} (1998) 683, {\tt hep-th/9806016}.

\bibitem{Witten92}
E. Witten, `On holomorphic factorization of WZW and coset models', {\it Commun. Math. Phys.} {\bf 144} (1992) 189.

\bibitem{Griffiths-Harris}
P. Griffiths and J. Harris, {\it Principles of Algebraic Geometry}, (Wiley 1978).

\bibitem{Polchinski}
J. Polchinski, `Evaluation of the one-loop string path integral', {\it Commun. Math. Phys.} {\bf 104} (1986) 37.

\end{thebibliography}
\end{document}